\documentclass[aps,prd,letterpaper,11pt,twoside,tightenlines,nofootinbib,showpacs,preprint,twocolumn]{revtex4}
\usepackage{graphicx}
\usepackage[sort&compress]{natbib}
\usepackage{latexsym}
\usepackage{epsfig}
\usepackage{amsmath}

\def\HH{\mathsf{H}} 

\def\be{\begin{equation}}
\def\ee{\end{equation}}

\begin{document}

\title{Gluon Shadowing and Nuclear Entanglement}
\author{Paolo Castorina$^{\rm 1,2,3}$, Alfredo Iorio$^{\rm 2}$, Daniele Lanteri$^{\rm 1,2,4}$ and Petr Luke\v{s}$^{\rm 2}$}
\affiliation{
\mbox{${}^1$ INFN, Sezione di Catania, I-95123 Catania, Italy} \\
\mbox{${}^2$ Faculty of Mathematics and Physics, Charles University} \\
\mbox{V Hole\v{s}ovi\v{c}k\'ach 2, 18000 Prague 8, Czech Republic}\\
\mbox{${}^3$} School of Nuclear Science and Technology, Lanzhou University, 222 South Tianshui Road, Lanzhou 730000, China\\
\mbox{${}^4$ Dipartimento di Fisica e Astronomia, Universit\`a  di Catania, I-95123 Catania, Italy.}}

\date{\today}
\begin{abstract}
Relying on previous results that link entanglement entropy and parton distribution functions in deep inelastic scattering and focusing on the small Bjorken scaling region we present here indications that  gluon shadowing might indeed be explained as due to a depletion of the entanglement entropy between observed and unobserved degrees of freedom per  nucleon within a nucleus with respect to the free nucleon case. We apply to gluon shadowing the general Page approach to the calculation of the entanglement entropy in bipartite systems, giving physical motivations of the results.
\end{abstract}
 \pacs{13.60.Hb,12.38.Cy,03.65.Ud}
 \maketitle

\section*{Introduction}

Quantum-Chromodynamics (QCD) evolution equations of  parton distribution functions (pdfs) in deep inelastic scattering (DIS) require specific initial conditions, i.e. the evaluation of the pdfs as a function of the Bjorken scaling variable, $x$, at a fixed  transfer momentum, $-q^2_0=Q^2_0$.

Ideed, the phenomenological analyses of pdfs~\cite{EPPS16,ncteq15} are crucial for a reliable comparison of QCD predictions with the experimental data in high energy collisions, taking into account the effective uncertainty in the evolution equations.

Moreover, a quantitative description of the modifications of pdfs in nuclear environment is important to disentangle the so-called \textit{cold nuclear effects} from the genuine signatures of the formation of a quark-gluon plasma (QGP) at finite temperature and baryon density in relativistic heavy ion collisions. Indeed, the well known phenomenon of ``nuclear shadowing''~\cite{Armesto2006Review}, i.e. the depletion of pdfs per nucleon \textit{within} a nucleus  with respect to the  \textit{free} nucleon case in the small $x$ region can mimic the suppression of particle production due to the formation of the QGP.

The functional form of the initial pdfs is a nonperturbative problem related to quark and gluon confinement and a recent proposal~\cite{dima1} suggests that the gluon distribution function in the low $x$ region can be directly related to the entanglement entropy of the parton in the nucleon wave function. More precisely, the nucleon in its rest frame is described by a pure quantum mechanical state $|\psi>$ with density matrix $\rho=|\psi><\psi|$, such that the entanglement entropy
\be
S = - {\rm Tr} \left( \rho \ln \rho \right) \,,
\ee
is identically zero. However, DIS  probes only part of the nucleon's wave function, call it the spatial region $I$.

The inclusive DIS measurement thus sums over the unobserved part of the wave function localized in the region $II$, complementary to $I$, so one has access only to the \textit{reduced} density matrix, $\rho_I =  {\rm Tr}_{II} \rho$.

In~\cite{dima1} it turns out that in the small $x$ region ($x \leq 0.008$) the entanglement entropy $S_I \equiv - {\rm Tr} \left( \rho_I \ln \rho_I \right)$ is related to the gluon distribution $xG(x)$ as follows
\begin{equation}
\begin{split}\label{eq:2}
S_I \equiv & S_N(x) = \ln[xG(x)]  \\
= & \Delta \; \ln[1/x] = \Delta \; \ln[L/\epsilon] \,,
\end{split}
\end{equation}
where $L=1/(m_N x)$, with $m_N$ the nucleon's mass, $\epsilon =1/m_N$ is the Compton wavelength of the nucleon and  $\Delta=0.15$, for the free nucleon~\cite{altri} at $Q_0^2 \simeq 1.7$ Gev$^2$.

Taking this point of view, gluon shadowing should be due to a \textit{reduction} of the entanglement between observed and unobserved degrees of freedom (d.o.f.) for a \textit{bounded} nucleon with respect to a \textit{free} nucleon.

At a first glance, this conclusion could appear to contradict the idea that the entanglement entropy, $S_I$, for a subsystem $I$ of fixed (quantum) dimension $m$ is bigger when the combined system $I \bigcup II$ has a bigger (quantum) dimension $N = m n$.

However, the crucial aspect is which d.o.f. are \textit{effectively} entangled in any given specific process. It is the intent of this paper to propose a model-independent analysis of gluon shadowing based on entanglement entropy where a dynamical mechanism selects the effective entangled d.o.f. and  produces  a \textit{smaller} $S_I$.

In Section~\ref{sec:1} we recall the general argument by Page~\cite{pagebipartite,pageBH} on the average entropy of a quantum subsystem, based on the unitary evolution of a pure state into a pure state. Section~\ref{sec:2} is devoted to the application of Page's formulas of entanglement entropy to gluon shadowing. In Sec.~\ref{sec:3} we discuss some physical motivations of the obtained results and Sec.~\ref{sec:cc} is devoted to comments and conclusions.

\section{\label{sec:1} Entanglement entropy and Page curves}

Consider the Hilbert space $\HH$ of a generic quantum bipartite system, $\HH = \HH_I^m \otimes \HH_{II}^n$,
where $m$ and $n$ indicate the dimension, so that $\dim \HH = m\,n \equiv N$. Pick up there an arbitrary state
$|\psi_0> \in \HH$, and a random unitary matrix $U$, so that $U |\psi_0>$ is a random state in $\HH$. If we trace out the subsystem $II$, we associate the density matrix $\rho_I(U)$ to this state and thus also the related entanglement entropy $S_I(U)$. We can then average through $U$ to get the average entanglement entropy \footnote{A related important quantity is the average \textit{information}, defined as $i \equiv \ln m - S_I$.} of the subsystem $I$,
\begin{align}\label{eq:3}
  S_I &= \left\langle S_I (U) \right\rangle_{\text{average through $U$}} \, .
\end{align}
Page conjectured, as was later proven by Sen \cite{sen}, that
\begin{eqnarray}
      & & \sum_{k=n+1}^{N} \frac{1}{k} - \frac{m-1}{2n}, \qquad \text{for}~m \le n \, , \label{firsthalf}\\
  S_I &=&  \nonumber \\
      & & \sum_{k=m+1}^{N} \frac{1}{k} - \frac{n-1}{2m}, \qquad \text{for}~m \ge n \, . \label{secondhalf}
\end{eqnarray}
Notice the symmetry under $n \leftrightarrow m$ of the first and second expression. This is a manifestation of the general result, $S_I = S_{II}$, valid for an overall pure system. Here this means that, when the observable d.o.f., $m$, become bigger than the unobservable d.o.f., $n$, then one can reverse the point of view, and use $S_{II}$ instead.

Some of the mathematical details of this construction can be seen in~\cite{pageBH}, where this was applied to black hole (BH) evaporation (see also the review~\cite{HarlowReview}). In that application, it is assumed that BH evaporation takes an initial pure state into a final pure state, while middle states are mixed. The total Hilbert space is factorized into a product, where the subsystem $I$ corresponds to the \textit{unobserved} states under the horizon, and the subsystem $II$ corresponds to the \textit{observed} Hawking radiation. When the BH is formed, there is no radiation outside and, hence, $m = 1$ and $n = N$, thus, $S_I$ is trivially zero (pure state). As the BH evaporates, $N = n m$ is constant, but $m$ increases while $n$ decreases, and, according to~\eqref{firsthalf}, $S_I$ increases. Approximately at half time of the evaporation the two subsystems have the same quantum dimensions, $n = m$, the information stored below the horizon starts to leak from the BH decreasing $S_I$, according to~\eqref{secondhalf}. When the BH fully evaporates, $n = 1$ and $m = N$ and $S_I$ returns to zero (pure state).

\begin{figure}
	\centering
	\includegraphics[width=0.5\textwidth]{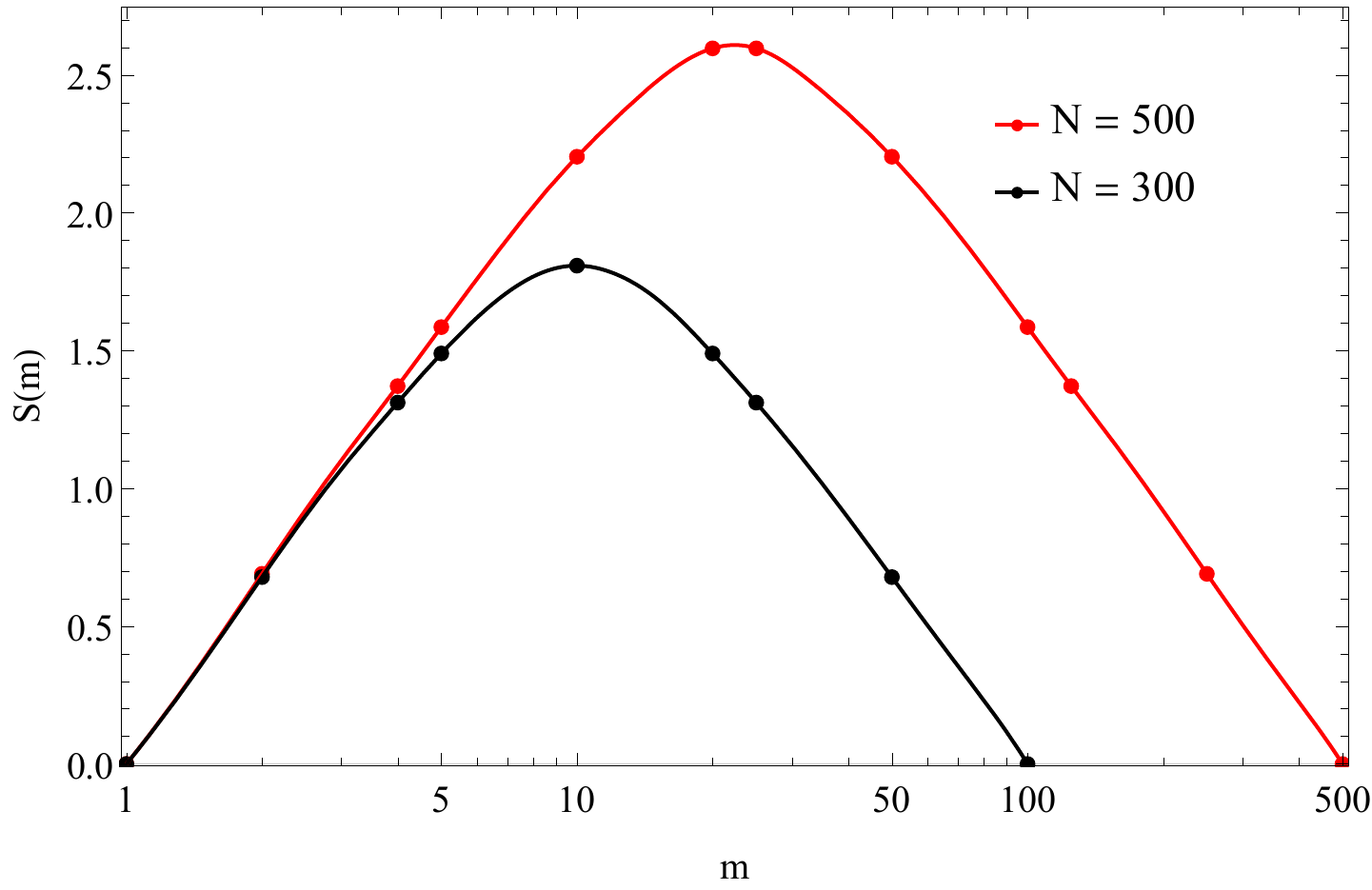}
	\caption{Example of two Page's curves with total numbers of degrees of freedom 500 and 300. The points are interpolated by spline. Horizontal axis describes entanglement entropy, vertical axis shows the number of degrees of freedom in one part of the bipartite system.}
	\label{fig:two_pages}
\end{figure}

In Fig.~\ref{fig:two_pages} we plot two such processes for two different $N$s. The figure shows that, for a fixed dimension of the observable subsystem, $m$, $S_I$ is bigger for bigger $N$, hence bigger $n$, the unobservable d.o.f. In our application here to nuclear shadowing, we propose that the depth of the region probed by the DIS, that is inversely proportional to $x$, is related to $m$: the bigger the region probed, the more $m$ are within reach. A full analysis of how this curve might apply to the various regimes of shadowing and anti-shadowing~\cite{Armesto2006Review} is beyond the scope of this paper, as we focus on the small $x$ region, and no matter which part of the curve is relevant, what we need to know is that the curve for $N'$ is always above the curve for $N$ when $N' > N$.

Before moving ahead, let us close this Section by mentioning that different results are obtained for the evolution of a bipartite quantum system such as a BH, when one allows for quasi-particles states (that is non-pure states) as ending states, as shown in~\cite{beyondPage}.

\section{\label{sec:2} Gluon Shadowing \`a la Page}

According to the previous Section, for any given $n$, we have $S'_I(n) > S_I(n)$ when the total dimension of the primed bipartite system, $N' = m' n$, is bigger than the total dimension of the unprimed bipartite system, $N = m n$. I.e., when, for any given $n$, it is always $m' > m$.

Let us now recall that for a system of $K$ independent nucleons the global gluon pdf is the product of the free gluon distribution of the single nucleons, i.e.
\be
(xG)^{free}_K = \prod_{i=1}^K (xG)_i^{free}
\ee
and therefore the corresponding  entanglement entropy is given by
\be
S_K^{free} = K S_N.
\ee
To find the gluon pdf for a nucleus of atomic number $A$, $(xG)^A$, is a nonperturbative and difficult dynamical problem. In this paper a simplified approach is proposed, where a modification of the free nucleon pdf takes into account the nuclear effects. In other words, following Eq.(6), one considers the factorization
\be
(xG)^A = \prod_{i=1}^K (xG)_i^{N/A}
\ee
where $ (xG)^{N/A}$ is the modified gluon pdf of the single bounded nucleon.
Therefore for the nuclear entanglement entropy one obtains
\be
S_A = A \ln [ (xG)^{N/A}]
\ee
and $S_A/A$ is the entanglement entropy per nucleon.

The shadowing effect can now be described in terms of entanglement entropy, since the ratio between the gluon pdfs in nuclei with atomic numbers $A$ and $B$ turns out to be
\be
\exp\left(\frac{S_A}{A} - \frac{S_B}{B}\right) = \frac{xG^{N/A}}{xG^{N/B}}
\ee

This simple model is supported by the following argument on the entanglement of a bipartite systems, when the two parts are modified in different ways by dynamical effects. For a free nucleon, the d.o.f. on the two sides of the separation $I$ and $II$ are all of the same kind. Whether truly fundamental or not, let us call them ``fundamental'', as they refer to the simplest nucleon system. Henceforth, each such ``fundamental'' d.o.f in the observed region $I$ can be entangled at most with one d.o.f. of the unobserved region, $II$, according to the standard monogamy of entanglement. Therefore, depending on how the d.o.f. are shared between $I$ and $II$, we might have partial or full (maximal) entanglement. Now, suppose that, \textit{within the nucleus}, the d.o.f. of $I$ and the d.o.f. of $II$ are affected by the highly nontrivial dynamical effects. Such effects then rearrange the d.o.f. inside the two regions and clearly the pairing between d.o.f. on the two sides is affected too. One might think of such rearrangement as a map from the ``fundamental'' d.o.f. of the bounded nucleon to the modified d.o.f, of the ``quasi-particles''. The latter is the picture behind the model presented above.

Let us now discuss the consequences of Eq.(10).

If $A>B$, only a reduction of the effective entangled d.o.f. from a larger nucleus to a smaller one can explain the gluon shadowing. Some dynamical mechanisms producing this depletion will be discussed later and in this section we strictly apply Page's formulas to describe gluon shadowing.

Equations (\ref{firsthalf},~\ref{secondhalf}) can be written in the form
\begin{equation}
S=
\begin{cases}
\displaystyle \sum_{k=1}^N\frac{1}{k} - \frac{m-1}{2\;n}  - \sum_{k=1}^n\frac{1}{k}& m\leq n\\
\displaystyle \sum_{k=1}^N\frac{1}{k} - \frac{n-1}{2\;m} - \sum_{k=1}^m\frac{1}{k}  & m\geq n\\
\end{cases}
\;.
\end{equation}
Since for large $N$
\be
\sum_{k=1}^N\frac{1}{k} \simeq \ln(N) + \gamma_M,
\ee
where $\gamma_M$ is the Eulero-Mascheroni constant, it turns out that
\begin{equation}
S
=
\begin{cases}
\displaystyle \ln m - \frac{m(m-1)}{2\;N}  \qquad m\leq \sqrt{N}\\\\
\displaystyle \ln\frac{N}{m} -\frac{N-m}{2\;m^2}\qquad m> \sqrt{N}
\end{cases}
\end{equation}
and
\begin{equation}
e^{S}
=
\begin{cases}
\displaystyle m\;\exp\left\{-\frac{m\;(m-1)}{2\;N}\right\}\qquad m\leq \sqrt{N}\\\\
\displaystyle \frac{N}{m}\;\exp\left\{-\frac{N-m}{2\;m^2}\right\}\qquad m> \sqrt{N}\\
\end{cases}
\end{equation}
The result in eq.~\eqref{eq:2}  has been obtained by considering a bipartite system
(observed and unobserved part of the free nucleon wave function).
From this point of view Page's general formulas can be applied, giving a relation among the d.o.f. $m$ (or $n$), $N$ and the free nucleon gluon pdf. Indeed, for $m < \sqrt{N}$, one has
\be\label{eq:11}
S_D = \ln m_D - \frac{m_D(m_D-1)}{2\;N_D} = \ln [xG_D(x)]
\ee
where $S_D$ and $xG_D$ are respectively  the entanglement entropy  and the gluon pdf per nucleon in deuterium, which we identify with the free nucleon ones, neglecting the small binding effects.

In perturbative QCD $xG(x, Q)$ obeys the BFKL evolution equation~\cite{bfkl} and grows at small $x$ as $(1/x)^\Delta$, with $\Delta \simeq 0.15$ \cite{altri} at $Q_0^2 \simeq 1.7$ Gev$^2$, and therefore (see eq.~\eqref{eq:11})
\be\label{eq:12}
S_D = \ln m_D - \frac{m_D(m_D-1)}{2\;N_D} = \ln\left[\left(\frac{c}{x}\right)^\Delta\right]
\ee
where $c=3.34$ is a normalization constant fixed by imposing the phenomenological value of $xG_D(x) \simeq 5$ at $x=10^{-4}$~ ( see tab.3 of ref. \cite{value} and ref. \cite{th} ).

\begin{figure}
  \centering
  \includegraphics[width=0.5\textwidth]{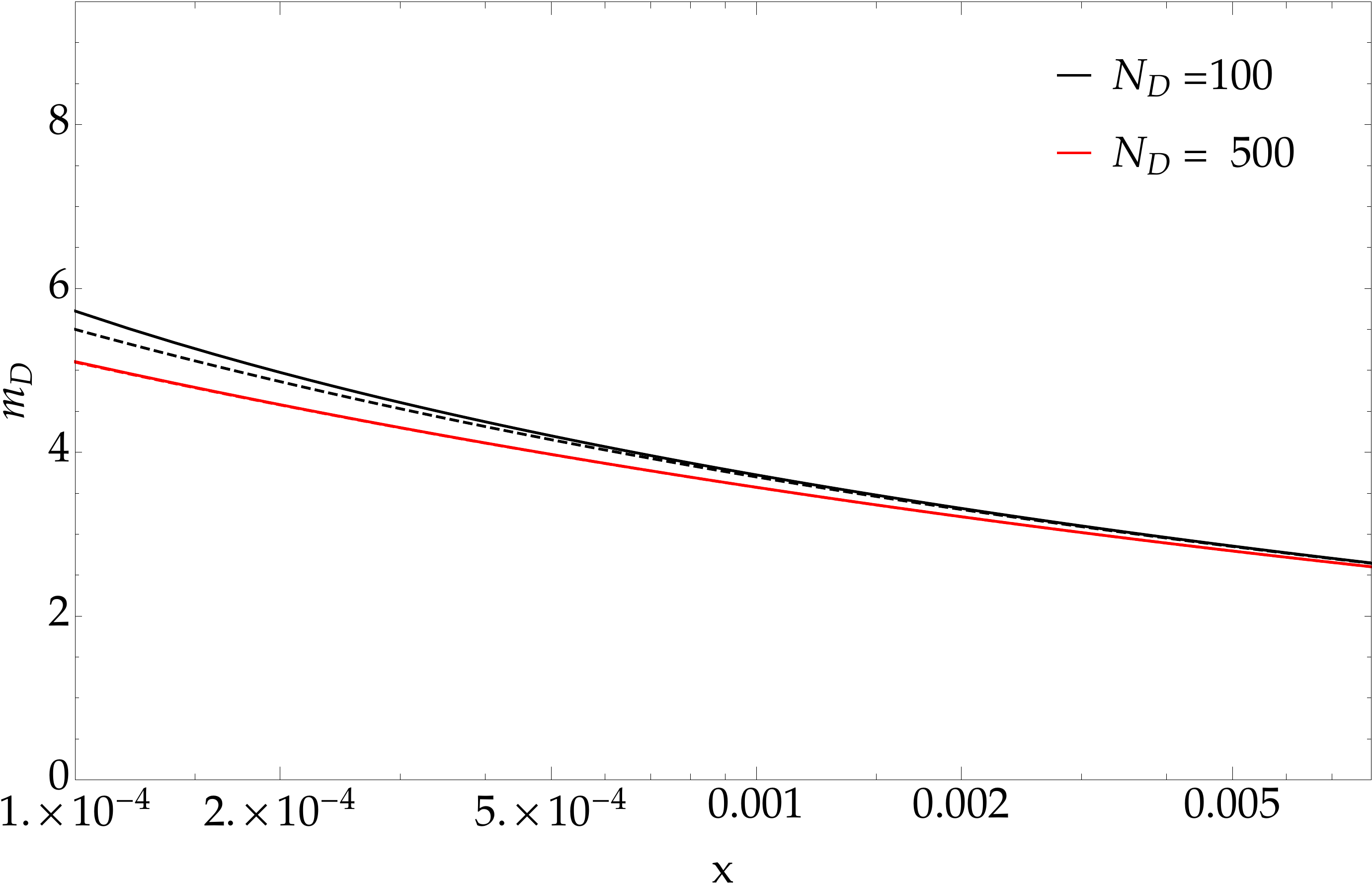}
  \caption{$m_D$ as a function of  $x$ from eq.~\eqref{eq:12} compared with the approximation in eq.~\eqref{eq:13} for different values of $N_D=m_D n_D$}
  \label{figuno}
\end{figure}

By previous equation one gets the correpondence between the d.o.f. $m_D$ and the Bj variable $x$ for a free nucleon depicted in fig.~\ref{figuno}, which for
$N_D >> 1$ can be approximated by
\be\label{eq:13}
m_D(x) = \left(\frac{c}{x}\right)^\Delta \left[ 1 + \frac{1}{2\,N_D} \left(\frac{c}{x}\right)^{\Delta}\left(\left(\frac{c}{x}\right)^{\Delta}-1\right)\right]
\,.
\ee

Also for a nuclear target, $A$, DIS  probes only part of the {\it bounded} nucleon's wave function and therefore
the description of shadowing in terms of a different entanglement entropy for a bounded nucleon with respect to the free nucleon case requires that $m_A$, the d.o.f. per nucleon in $A$ in  eq.~\eqref{eq:11} compared with $m_D$ in eq.~\eqref{eq:12} has to be such that  $S_A < S_D$ in the small $x$ region.

\begin{figure}
  \centering
  \includegraphics[width=0.5\textwidth]{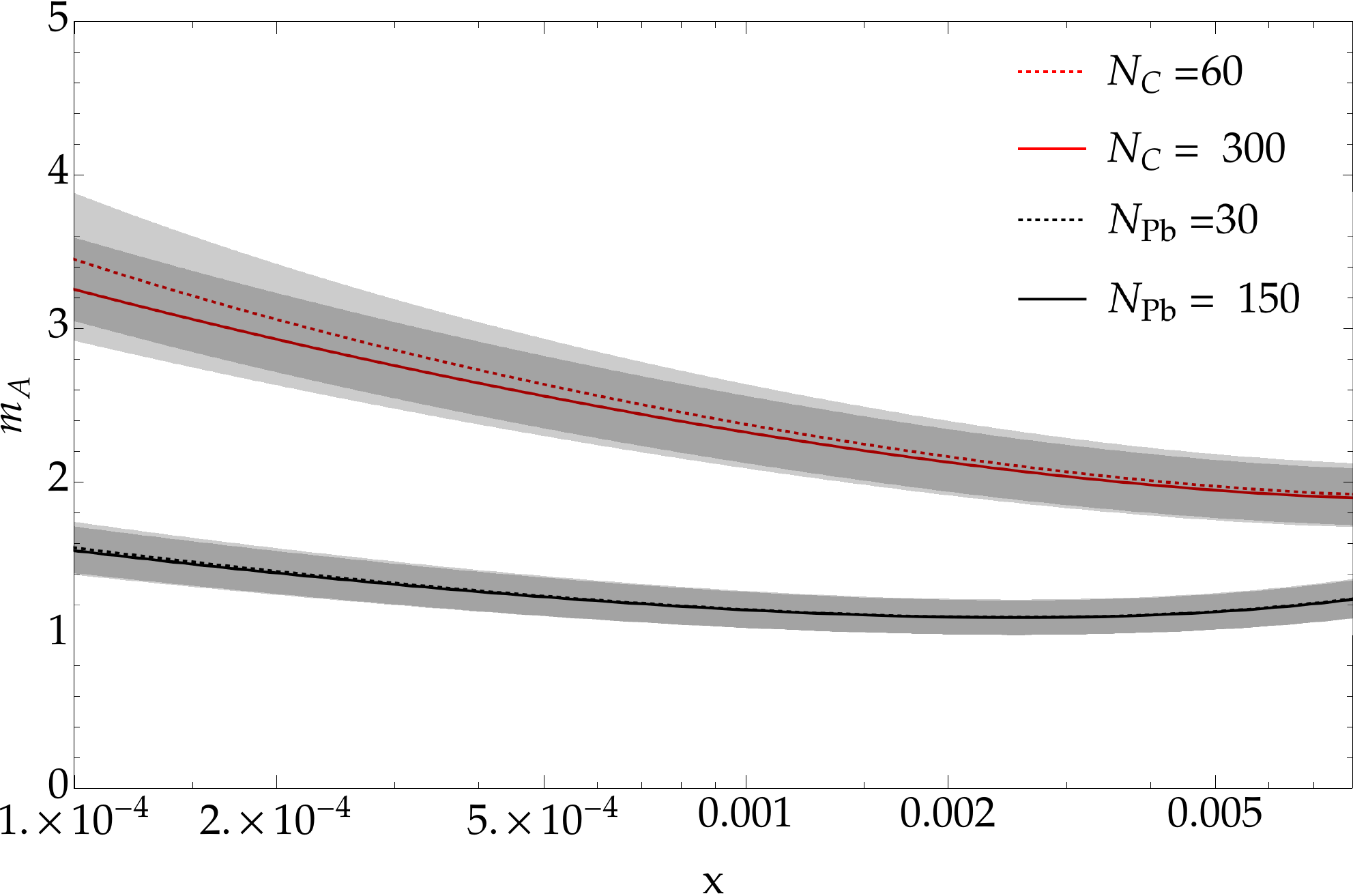}
  \caption{$m_A(x)$ that fits the phenomenological results on gluon shadowing for $x < 0.008$~\cite{EPPS16}  for two different nuclei (C and Pb) and different values of $N_C$ and $N_C$. The grey bands are due to
the corresponding uncertainties in the gluon distribution function.}
  \label{figdue}
\end{figure}

\begin{figure}
  \centering
  \includegraphics[width=0.5\textwidth]{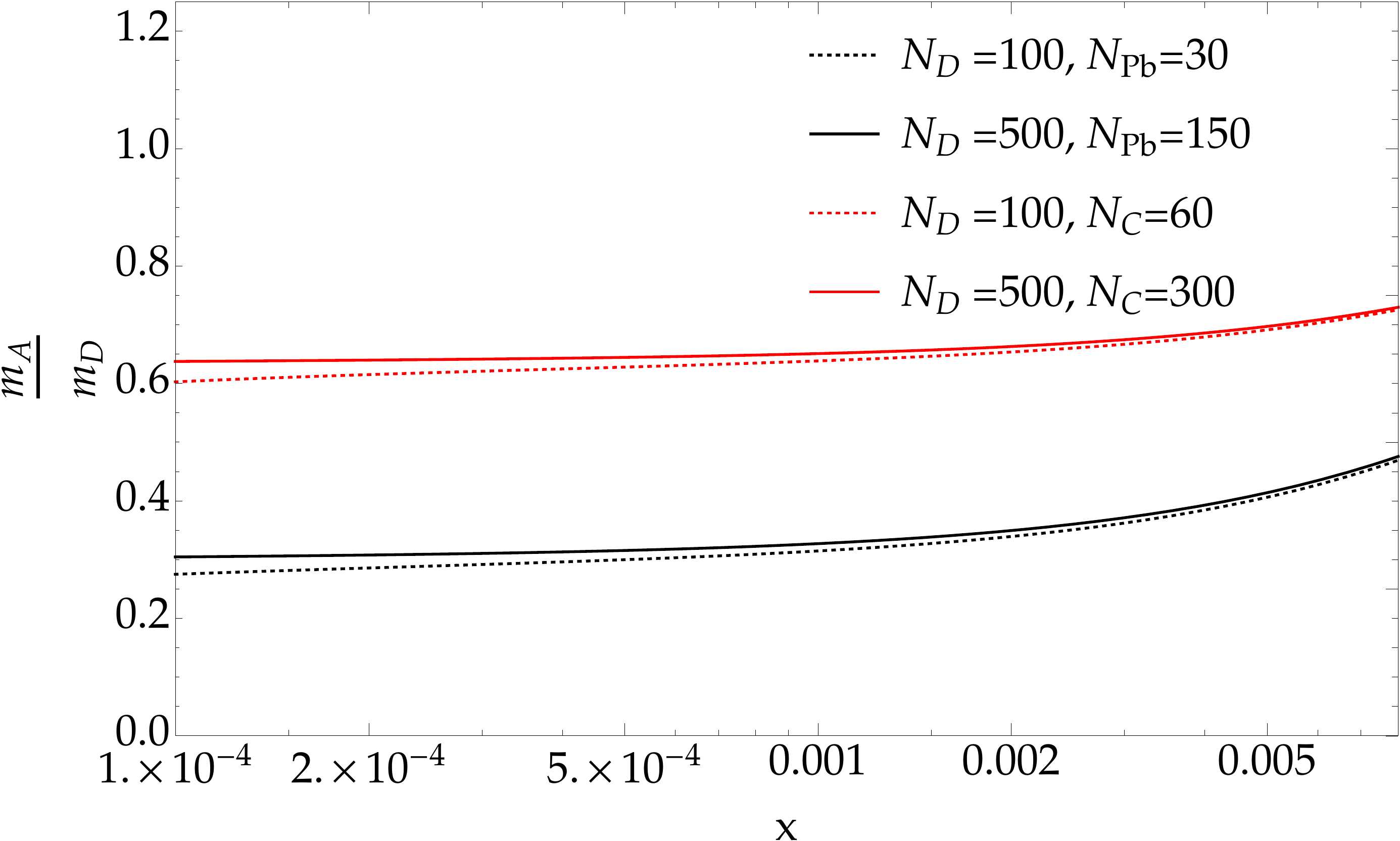}
  \caption{Ratio $m_A/m_D$ as a function of $x$ for C and Pb and different value od $N_D,N_C,N_{Pb}$}
  \label{figtre}
\end{figure}

In fig.~\ref{figdue} the functions $m_A(x)$ that fits the phenomenological results on gluon shadowing for $x < 0.008$ at $Q_0^2 \simeq 1.7$ Gev$^2$ reported in ref.~\cite{EPPS16}  for two different nuclei (C and Pb) are shown for $N_D,N_A >>1$ and in fig.~\ref{figtre} there is depicted the corresponding ratio $m_A/m_D$.

\begin{figure}
  \centering
  \includegraphics[width=0.5\textwidth]{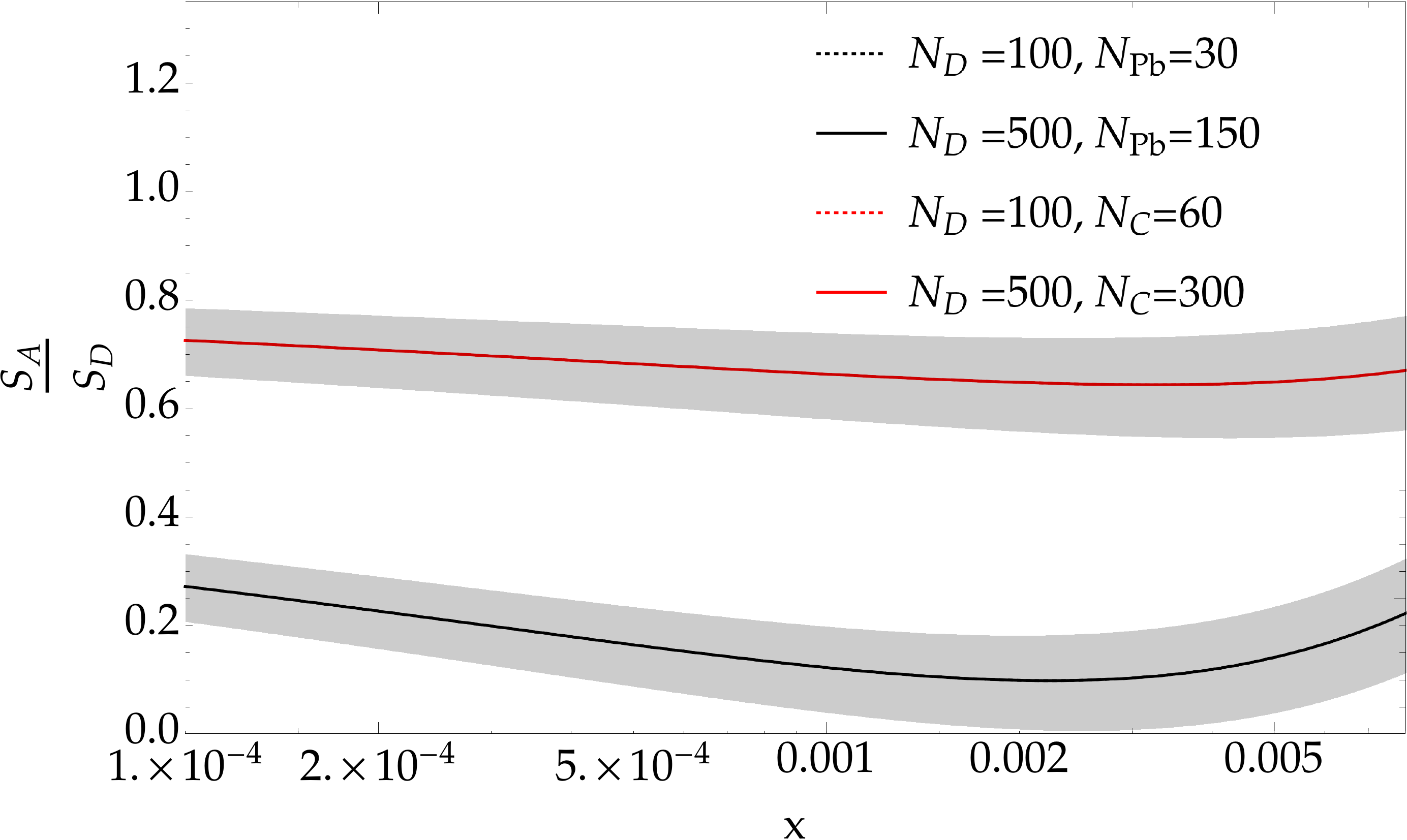}
  \caption{Ratio $S_A/S_D$ as a function of $x$ for C and Pb and different value od $N_D,N_C,N_{Pb}$.The grey bands are due to
the corresponding uncertainties in the gluon distribution function.}
  \label{figquattro}
\end{figure}

The previous analysis clearly shows that:
\begin{enumerate}
	\item the number of d.o.f. per nucleon probed by DIS in a nucleus is less than the corresponding d.o.f. for a free nucleon, despite the fact that one should expect a larger dimension of the whole Hilbert~space for a nucleus.
	
	\item by the results of $m_A(x)$ and $m_D(x)$ the ratio $S_A/S_D$ plotted in fig.~\ref{figquattro}, turns out to be almost constant in a large range of (small) $x$, with $S_{PB}/S_D \simeq 0.3$ and
	$S_{C}/S_D \simeq 0.7$.
\end{enumerate}

\section{\label{sec:3} Dynamical mechanics for nuclear entanglement}

The results in Sec.2 are based on the general Page's framework and  points (1-2) call for  possible dynamical interpretations.

A possible, simple mechanism is related to the overlap of nucleons in the nucleus~\cite{bob}: due to the larger overlap in nuclei of atomic number $A$ and $B$, with $A>B$, the observed number of d.o.f. per nucleon decreases which in general implies  that unobserved d.o.f. per nucleon increases.
By assuming the dominance of two nucleons overlap in nuclei~\cite{bob} and defining $V_0^A$ the overlap per nucleon in nucleus $A$,
for the ratio $S_A/S_D$ one expects
\be
S_A/S_D = 1-V_0^A
\ee
The overlapping volume per nucleon has been evaluated in ref.~\cite{bob}  (table I, column b)  and for the second member of the previous ratio one gets
$\simeq 0.28$ for $Pb$ and $\simeq 0.6$ for $C$ which  are  compatible with the ratios in fig.~\ref{figquattro} if one takes into account the large uncertainty band in the gluon pdf.

Another dynamical interpretation of the depletion of d.o.f per bounded nucleon  comes from ref.~\cite{cornwall} where it has been suggested that the gluon condensate  in vacuum, at zero temperature,
\be
<{\alpha_s \over \pi} G_{\mu \nu}^a (0) G^{\mu \nu a} (0)>
\ee
has a large entropy, arising from lumps of condensate with a finite correlation length. On the other hand, the gluon condensate shrinks when the baryon density increases~ \cite{altricond,bcz} and, despite  this result referring to nuclear matter, one can expect this depletion to come out, for instance, by considering light and heavy nuclei.

In fact, this depletion of the gluon condensate and consequently of the related entropy has a role in the DIS structure functions.
In the small $x$ region the structure function is described by the pomeron exchange~\cite{tredici}
and for $x \rightarrow 0$ the structure functions are related
to the quark-pomeron effective coupling in the target (T), $\beta_T$,
\be
\label{eq:15}
F_2(x) _ {\longrightarrow\atop {x \rightarrow 0}} \; \beta_T,
\ee
and the ratio of the structure functions per nucleon in the nucleus A with respect to nucleus B is given by~\cite{quattordici}
\be
\label{eq:16}
{{{F_2(x)^A \over {F_2(x)^B}}} \vert _  {\longrightarrow\atop {x \rightarrow 0}} \; {\beta_A \over {\beta_B}}}
\ee

On the other hand, in the model proposed by Landshoff and Nachtmann ~\cite{quindici} the pomeron exchange is described by
nonperturbative gluons and  the quark pomeron coupling is related to the gluon condensate according to~\cite{sedici,diciasette}
\be
\label{eq:17}
\beta_T \simeq a^5 < T |{\alpha_s \over \pi} G_{\mu \nu}^a (0) G^{\mu \nu a} (0)| T >.
\ee

One can consider the correlation length $a$ as independent of the target, because
it is connected to the pomeron coupling to off-shell quarks~\cite{sedici,diciasette} and therefore one obtains
\be
\label{eq:18}
{{{F_2(x)^A \over {F_2(x)^B}}} \vert _  {\longrightarrow\atop {x \rightarrow 0}} \; {\beta_A \over {\beta_B}}}
= { { < A |{\alpha_s \over \pi} G_{\mu \nu}^a (0) G^{\mu \nu a} (0)| A >} \over
{< B |{\alpha_s \over \pi} G_{\mu \nu}^a (0) G^{\mu \nu a} (0)| B >}}
\ee
which suggests the reason for the depletion of entropy in DIS at small $x$ through the nuclear modification of the gluon condensate.

Another interesting aspect is the origin of antishadowing. A dynamical model of antishadowing of the  gluon distribution has been proposed in ref.~\cite{strick}, although
antishadowing is usually obtained in phenomenological analyses by imposing the energy-momentum sum rule. In the entanglement interpretation of nuclear pdf per nucleon, antishadowing should be related to
the conservation of the total entropy, i.e.
\be
\int_0^1 dx S_A(x) = \int_0^1 dx S_D(x)
\ee
which would require a complete description of parton pdf in terms of entanglement entropy in the whole $x$ range. In this respect the calculation of entropy in the valence approximation~\cite{david}, i.e.
for quarks in the color singlet baryon state, gives results consistent with the previous ones, since $S_B = \ln 3 \simeq 1.1$ for $x \rightarrow 1$ for a free nucleon (see eq.~(3.4) of ref.~\cite{david}) and, according to the previous normalization, $S_D \simeq 1.6$ for $x= 10^{-4}$.

\section{\label{sec:cc} Comments and Conclusions}

We put next to each other the evolution of the entanglement entropy of a bipartite system and the phenomenon of the nuclear shadowing. The first phenomenon is behind the famous Page curve of black-hole evaporation, in the unitary picture~\cite{pagebipartite,pageBH}, while the latter nuclear phenomenon refers to the depletion of the parton distribution functions per nucleon \textit{within} a nucleus, with respect to the \textit{free} nucleon. Relying on~\cite{dima1}, where entanglement entropy and parton distribution functions are found to be related, we wanted to move here the first steps towards understanding gluon shadowing in that approach. We leave it for future work to use more refined versions of Page curves~\cite{beyondPage}.

At first sight, the Page curves point to a behavior that clashes with shadowing as the entanglement entropy increases for larger unobserved degrees of freedom (i.e., larger nuclei). On the other hand, we have focused on the small $x$ region and we have presented indications that the shadowing might indeed be explained as due to a depletion induced by the overlapping volume per nucleon in a nucleus: the larger the overlap (hence, the larger the nucleus), the less degrees of freedom entangle.

Of course the proposed approach is just indicative and no detailed description of gluon shadowing could be obtained at this stage. Nonetheless, we are able to successfully compare our formulae with the EPPS16 fit of gluon shadowing for C and Pb. This shows that the entanglement interpretation of the depletion of gluon distribution in nuclei is a correct first step towards a more detailed dynamical description.

\section*{Acknowledgements}

A.I. is partially supported by UNCE/SCI/013.

\end{document}